\PassOptionsToPackage{unicode,pdfusetitle,bookmarksnumbered}{hyperref}
\RequirePackage{hyperref}
\documentclass[sigconf]{acmart}
\usepackage{etex}

\usepackage{subcaption} 

\let\terms\undefined

\acmBooktitle{}

\usepackage{booktabs}   
\usepackage{subcaption} 
\usepackage[noend]{algpseudocode}

\usepackage{adjustbox}

\usepackage{standalone} 

\usepackage{pgf-pie}

\usepackage{comment}

\usepackage{threeparttable}

\usepackage{graphicx}
\usepackage{amsmath}

\usepackage{xspace}
\usepackage{footnote}
\usepackage{wrapfig}  

\usepackage{amsfonts}
\usepackage{natbib}
\usepackage{hhline}
\usepackage{multirow}
\usepackage{setspace} 
\usepackage{url}
\usepackage{booktabs}
\usepackage{xcolor}
\usepackage{lipsum}
\usepackage{courier}
\usepackage{listings}
\usepackage{caption}
\usepackage{colortbl}
\usepackage{arydshln} 
\usepackage{makecell} 
\usepackage{tikz}
\usepackage{xcolor}
\usetikzlibrary{calc}
\usepackage{varwidth}
\usepackage{natbib}
\usepackage{algorithm}
\usepackage[noend]{algpseudocode}
\usepackage{subcaption}
\usepackage{wrapfig}

\usepackage[most]{tcolorbox}
\usepackage{tabularx}

\newcolumntype{Y}{>{\raggedright\arraybackslash}X}
\newcommand{\pp}{\nobreak\hspace{0.08em}pp\xspace}

\usetikzlibrary{fit,arrows,calc,positioning,quotes}
\usetikzlibrary{decorations.pathreplacing}

\usepackage{mathtools}
\usepackage{fancyvrb}
\usepackage{textcomp}
\usepackage{stmaryrd}

\newcommand{\app}{\texttt{EcoScratch}\xspace}

\tikzset{
  -|-/.style={
    to path={
      (\tikztostart) -| ($(\tikztostart)!#1!(\tikztotarget)$) |- (\tikztotarget)
      \tikztonodes
    }
  },
  -|-/.default=0.5,
  |-|/.style={
    to path={
      (\tikztostart) |- ($(\tikztostart)!#1!(\tikztotarget)$) -| (\tikztotarget)
      \tikztonodes
    }
  },
  |-|/.default=0.5
}

\usetikzlibrary{shapes.geometric}

\long\def\com#1{}

\newcommand{\para}[1]{\smallskip\noindent {\bf #1}}

\newcommand{\squishlist}{
   \begin{list}{$\bullet$}
    { \setlength{\itemsep}{0pt}      \setlength{\parsep}{3pt}
      \setlength{\topsep}{3pt}       \setlength{\partopsep}{0pt}
      \setlength{\leftmargin}{3.5mm} \setlength{\labelwidth}{1em}
      \setlength{\labelsep}{0.5em} } 
}

\newcommand{\squishend}{
    \end{list}  }

\def\BibTeX{{\rm B\kern-.05em{\sc i\kern-.025em b}\kern-.08em
    T\kern-.1667em\lower.7ex\hbox{E}\kern-.125emX}}

\author{Yuan Si}
\affiliation{%
  \institution{University of Waterloo}
  \city{Waterloo}
  \country{Canada}
  }
\email{yuan.si@uwaterloo.ca}

\author{Ming Wang}
\affiliation{%
  \institution{The Chinese University of Hong Kong}
  \country{Hong Kong}
  }
\email{cd.melvin.wang@gmail.com}

\author{Daming Li}
\affiliation{%
  \institution{Independent Researcher}
  \country{USA}
  }
\email{damingliyale22@gmail.com}

\author{Hanyuan Shi}
\affiliation{%
  \institution{Independent Researcher}
  \country{China}
  }
\email{shihanyuan1995@gmail.com}

\author{Jialu Zhang}
\authornote{Corresponding Author}
\affiliation{%
  \institution{University of Waterloo}
  \city{Waterloo}
  \country{Canada}
  }
\email{jialu.zhang@uwaterloo.ca}

\begin{document}

\begin{abstract}
	Scratch is the most popular programming environment for novices, with more than 1.15 billion projects created worldwide. Unlike traditional languages, correctness in Scratch is defined by visible behavior on the stage rather than by code structure alone, so programs that appear correct in the scripting workspace can still fail at runtime due to timing, event ordering, or cross-sprite interactions. Visual execution evidence such as gameplay videos can therefore be essential for diagnosis and repair. However, capturing and processing this evidence inside an automated repair loop introduces substantial overhead. Probing execution, recording stage behavior, rebuilding executable \texttt{.sb3} projects, and verifying candidate fixes consume time, monetary cost, and computational resources across an entire repair trajectory rather than a single model call.

We present \app, a repair pipeline that uses lightweight runtime signals to decide whether the next attempt stays text-only or escalates to multimodal prompting. The controller also sets the JSON Patch budget and verification effort, so evidence choice and repair budget are coupled inside the same decision. \app reconstructs candidate fixes into executable \texttt{.sb3} projects and records per-trajectory traces, monetary cost, and local-runtime energy.

We evaluate 12 models on 100 executable Scratch repair projects under four controller settings, yielding 4,800 repair trajectories. In this matrix, a selective multimodal policy gives the strongest observed success--cost--energy tradeoff. It reaches the highest generation success (30.3\%) while using less average cost and local-runtime energy than the two non-adaptive multimodal baselines under the same bounded trajectory budget; text-only remains the lowest-cost floor. Across the evaluated matrix, multimodal evidence helps most when it is used to control escalation within a bounded trajectory budget rather than applied uniformly.

\end{abstract}

\title{\app: Cost-Effective Multimodal Repair for Scratch Using Execution Feedback}

\maketitle

\section{Introduction}
\label{sec:intro}

Scratch is one of the most widely used programming environments in the world, with more than one billion shared projects and over 140 million children participating globally \cite{scratch2024}. At this scale, even small improvements in debugging support can affect millions of learner programs. Unlike traditional languages, correctness in Scratch is defined by visible behavior on the stage rather than by code structure alone. Animations, timing, sprite interactions, and broadcast ordering determine whether a program works. A project can appear locally reasonable in the scripting workspace yet fail completely once execution begins. Because Scratch is often a learner's first encounter with state, concurrency, and event-driven computation, debugging support directly shapes both success and how learners understand the connection between code and behavior.

Large language models make automated support for Scratch increasingly feasible, especially when they reason over execution evidence rather than static blocks alone. Many failures do not reside in a single script or local pattern. They emerge from cross-sprite interactions, event interleavings, timing mismatches, or discrepancies between intended and observed behavior. Prior work shows that visual runtime evidence such as screenshots or gameplay video can greatly improve diagnosis and repair \cite{si2025viscratch}. In principle, multimodal reasoning appears to be the natural solution for debugging behavior-driven programs.

In practice, however, multimodality introduces a severe deployment barrier. Capturing stage behavior, encoding visual inputs, transmitting them to a model, and processing them inside an automated repair loop can consume substantial time, monetary cost, and computational resources \cite{si2026scratcheval}. When scaled to real educational platforms where millions of programs may require assistance, always using multimodal reasoning becomes economically unsustainable and operationally impractical. Systems that rely on images or video for every case would be too slow, too expensive, and too resource-intensive to deploy continuously. The very evidence that improves repair accuracy can therefore prevent automated support from being used at all.

This creates a fundamental dilemma for executable repair systems. Stage-level evidence is sometimes essential because the root cause of a failure may be visible only during execution, yet using multimodal inputs indiscriminately can exhaust budgets before reaching many learners. A purely text-based system may remain blind to behavioral bugs, while an always-multimodal system may become unusable in practice. The central challenge is therefore not only how to repair programs, but how to decide when richer evidence is worth its cost.

In an executable repair pipeline, this decision unfolds across an entire probe, repair, and verification trajectory rather than a single prompt. Cost accumulates as the system probes the project, decides whether to capture images, constructs requests, generates candidate patches, rebuilds the program, executes the result, verifies behavior, and sometimes retries after malformed or ineffective outputs. Multimodal evidence is therefore not merely an accuracy enhancement. It is a resource allocation decision that affects monetary cost, execution overhead, and energy consumption across the whole trajectory. Model-side token counts alone do not reflect the true deployment budget.

This shifts the problem from model capability to repair-time control. We study executable Scratch repair as backend infrastructure that could operate behind learner-facing assistants or large-scale evaluation services. The target use case is a service that receives a failing executable project, can run bounded probe and verification loops, and returns a candidate repair under a fixed trajectory budget. We do not study tutoring dialogue, natural-language feedback quality, or open-ended critique of non-executable projects.

Existing systems demonstrate that automated Scratch support is feasible across static analysis, hint generation, debugging tools, and execution-grounded approaches \cite{morenoleon2015drscratch,fraser2021litterbox,fraedrich2021commonbugs,obermueller2021catnip,marwan2023isnap,deiner2024nuzzebug}. What remains unresolved is a deployment question that becomes central once repair is executable and multimodal evidence is available: how should the system allocate a bounded trajectory budget? Escalating too early wastes resources on easy programs, while remaining text-only for too long can leave the system unable to diagnose failures whose essential symptoms appear only on the stage.

We address this problem with \app, a runtime-signal-guided repair pipeline for Scratch. The pipeline begins with a lightweight execution probe that summarizes observable behavior and uses this signal as triage rather than full diagnosis. Instead of repairing immediately, the system selects a repair plan that determines whether to remain text-based or escalate to multimodal prompting, and jointly sets patch budgets and verification effort. Candidate fixes are expressed as JSON patches, reconstructed into executable \texttt{.sb3} files, and evaluated through staged verification. Each trajectory records request traces, patch outcomes, verification results, monetary cost, and local-runtime energy, which lets us analyze deployment tradeoffs rather than prompt-level accuracy alone.

We evaluate a complete $12 \times 4 \times 100$ matrix of models, controller modes, and executable Scratch repair projects. Our primary metric is generation success, defined as the proportion of trajectories that successfully produce a candidate patch. We also report strict verification success together with per-trajectory cost and local-runtime energy. These measures reveal not only whether repair succeeds, but how much evidence and system effort are required.

Our contributions are:

\begin{itemize}
    \item We present \app, an executable Scratch repair pipeline that couples runtime probing, plan selection, bounded patch generation, staged verification, and trajectory-level accounting.
    \item We conduct a deployment study across 12 models, four controller modes, and 100 executable repair projects under matched budgets.
    \item We show that selective multimodality achieves the strongest observed tradeoff, improving repair success while reducing both monetary cost and local-runtime energy relative to always using multimodal inputs.
\end{itemize}

\section{Background and Motivation}
\label{sec:background}

\begin{figure}[t!]
    \centering
    \includegraphics[width=.47\textwidth]{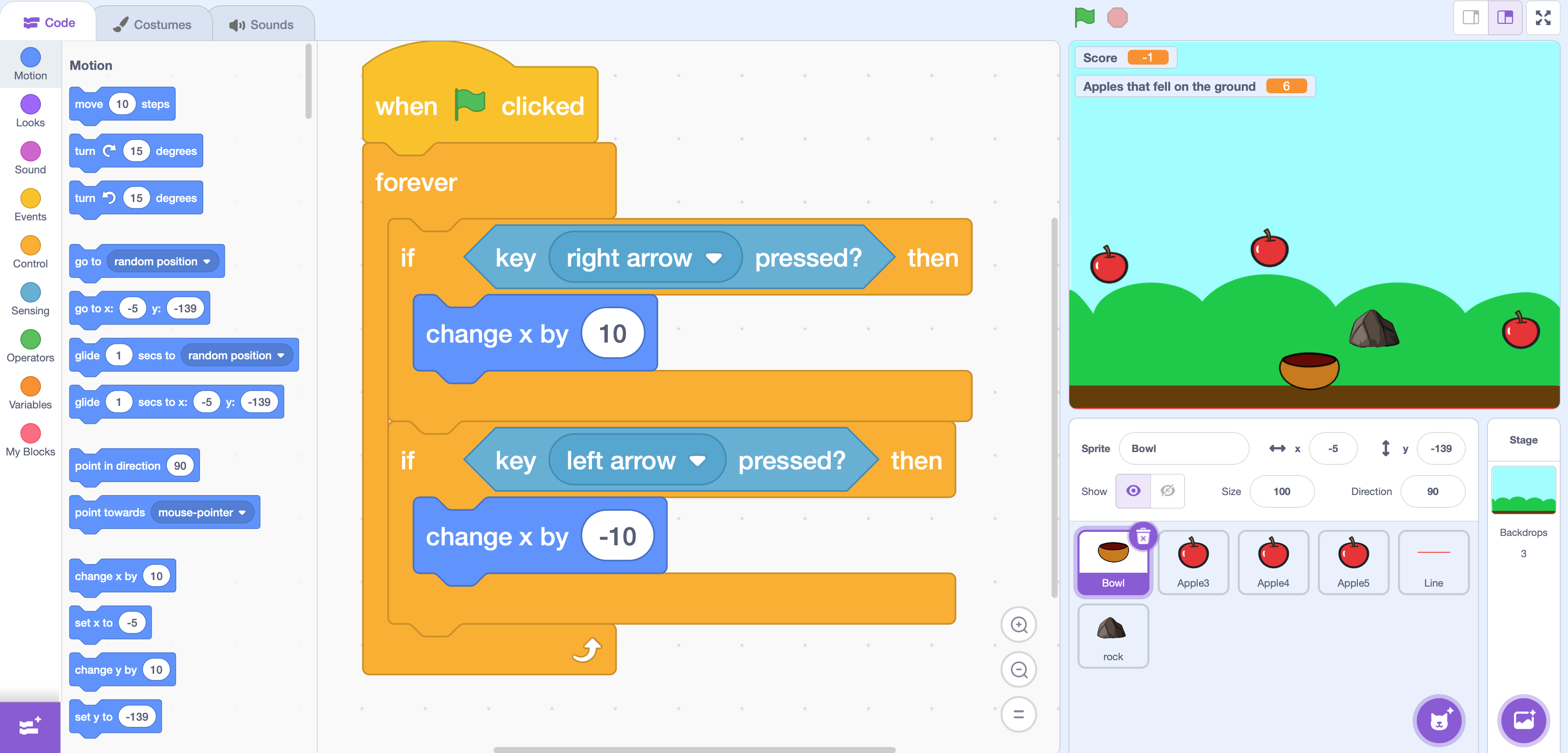}
    \caption{A Scratch project in the standard editor view. The scripting area exposes block structure, while the stage shows the behavior that ultimately determines whether the project works. The figure illustrates the mismatch studied in this paper: a code-local edit can look plausible in the scripting workspace yet still leave the visible runtime behavior wrong.}
    \label{fig:scratchui}
\end{figure}

\subsection{Scratch repair is a runtime-facing software engineering problem}

Scratch is programmed through blocks, but it is judged through behavior. Learners do not care whether a script looks tidy in isolation if the sprite still moves incorrectly, the broadcast never reaches the right handler, or the game never starts. What matters is whether the project behaves as intended on the stage. Figure~\ref{fig:scratchui} is the visual anchor for that split: the scripting workspace exposes block structure, while the stage shows the behavior users actually see. A typical failure looks locally plausible in code but obviously wrong on the stage: for example, a learner may wire what looks like the right green-flag or broadcast logic, yet the cat still never starts moving because the event reaches the wrong handler or a shared state change never occurs. Repair becomes hard when those two views diverge, because a code-local edit can look reasonable and still leave the visible failure untouched. This is the kind of case the controller studied later in the paper is designed for: a short probe can reveal that the visible symptom persists despite a plausible local edit, which is exactly when richer runtime evidence becomes worth collecting.

Scratch debugging is therefore not just local block correction. Scripts attached to different sprites may execute concurrently, broadcasts may trigger distributed behavior, and shared variables may couple logic across otherwise separate parts of a project. Scratch programs also mix code with media assets such as costumes and sounds \cite{resnick2009scratch,maloney2010scratch}, so failures often come from interactions among state, timing, assets, and visible outcomes rather than from a single malformed block. Prior work on common bugs, hint generation, adaptive feedback, and debugger support reports the same combination of structural causes, runtime interactions, and learner-visible symptoms \cite{fraedrich2021commonbugs,obermueller2021catnip,marwan2022aif,deiner2024nuzzebug}. Recent multimodal studies reach a similar conclusion: many Scratch failures become clear only when execution is observed directly \cite{si2025viscratch}.

A stronger generator does not remove that challenge. If the crucial symptom only appears at runtime, a better model can still miss it when it sees only block structure. Scratch therefore offers a compact but meaningful setting for a broader software-engineering question: how should an executable repair system decide what evidence to gather, when to escalate, and how much validation effort to spend? The projects are small, but the surrounding workflow already resembles the probe--escalate--verify pattern seen in executable evaluation and agentic repair systems \cite{stahlbauer2019testing,jimenez2024swebench,yang2024sweagent}.

\subsection{The cost of repair is incurred across the trajectory, not the model call}

A deployed repair system does not spend its budget at one isolated generation step. Scratch repair naturally follows a generate-and-validate loop: the system produces a candidate patch, applies it, executes it, and checks the result against the failure condition. This pattern is central to automated program repair \cite{legoues2012genprog,gazzola2019asr} and also appears in modern executable evaluation frameworks for Scratch and general software engineering \cite{deiner2023testgen,jimenez2024swebench,stahlbauer2019testing}. Once execution, rebuilding, and retries are inside the loop, a cheap model call can still belong to an expensive case. The meaningful budget unit is therefore the trajectory, not the first response.

Multimodal evidence makes the issue sharper. Collecting and transmitting images adds latency, bandwidth, and provider-side cost before the system even knows whether the case actually needs richer evidence. A policy that escalates too aggressively can overspend on cases that a text-only attempt would already solve, while a policy that stays text-only for too long can miss failures whose decisive symptoms appear only on the stage. Token totals alone do not capture this deployment reality. What matters is how much end-to-end work the repair trajectory required.

A Green AI perspective helps here. Green AI argues that systems should be evaluated through complete workflows rather than convenient proxies \cite{schwartz2020green}. For Scratch repair, that means counting the whole pipeline: probing, evidence preparation, request construction, patch generation, candidate reconstruction, verification, and retries. Token-only accounting can hide where resources are spent and can make a fixed multimodal policy look cheaper than it is on easy cases. The same argument applies to local-runtime energy, which is informative here only when measured over complete trajectories rather than isolated model calls.

\subsection{From multimodal capability to deployment control}

These observations reframe the problem. Prior work shows that visual evidence can improve Scratch debugging \cite{si2025viscratch}, but the unresolved question is when an executable repair system should pay to use that evidence under a bounded budget. In this setting, multimodality is not only a capability issue but also a control problem.

We therefore cast automated Scratch repair as a deployment control problem with three coupled decisions: whether to escalate from text-only to multimodal evidence, how much structural freedom to grant the patch generator, and how much verification effort remains worthwhile after each attempt. All three draw from the same limited budget, so optimizing them independently misses the real tradeoff. Always escalating may improve observability but overspend on easy cases, whereas conservative policies save cost but fail on runtime-visible bugs. The design task is to coordinate these choices at the trajectory level.

This framing guides our evaluation. We measure success, verification, image use, attempts, monetary cost, and local-runtime energy over complete repair trajectories rather than isolated model calls. 
Scratch is well suited to this systems question: projects are small enough for close inspection yet rich enough to expose real tradeoffs among evidence quality, controller policy, and deployment cost. Our goal is to determine when richer evidence is justified and how the controller should allocate that budget.

\section{System Design}
\label{sec:system}

\begin{table*}[t]
\caption{Operational glossary of the four controller modes. The controller mode is the outer evaluation condition; all four modes share the same runtime in Figure~\ref{fig:system_pipeline}, but differ in evidence policy and repair-plan selection.}
\label{tab:mode-glossary}
\centering
\footnotesize
\setlength{\tabcolsep}{5pt}
\renewcommand{\arraystretch}{1.15}
\begin{tabularx}{\textwidth}{
>{\raggedright\arraybackslash\bfseries}p{0.15\textwidth}
>{\raggedright\arraybackslash}p{0.30\textwidth}
>{\raggedright\arraybackslash}p{0.23\textwidth}
>{\raggedright\arraybackslash}X}
\toprule
Mode & Operational meaning & Evidence policy & Why it matters \\
\midrule
Text-only
& Never sends images; stays on the smallest text-only plan.
& No multimodal evidence.
& Lowest-cost and lowest-energy baseline. \\

Always-on multimodal
& Uses images whenever they are available on every case.
& Unconditional multimodal evidence.
& Isolates the cost of always escalating. \\

Fixed multimodal
& Applies one predetermined multimodal plan to every case.
& The same multimodal plan on every case.
& Separates a fixed multimodal policy from probe-guided scheduling. \\

Heuristic
& Uses runtime signals to stay text-only or escalate selectively.
& Multimodal evidence only when runtime signals justify it.
& Main proposed control policy. \\
\bottomrule
\end{tabularx}
\end{table*}

\begin{figure*}[t]
\centering
\includegraphics[width=\textwidth]{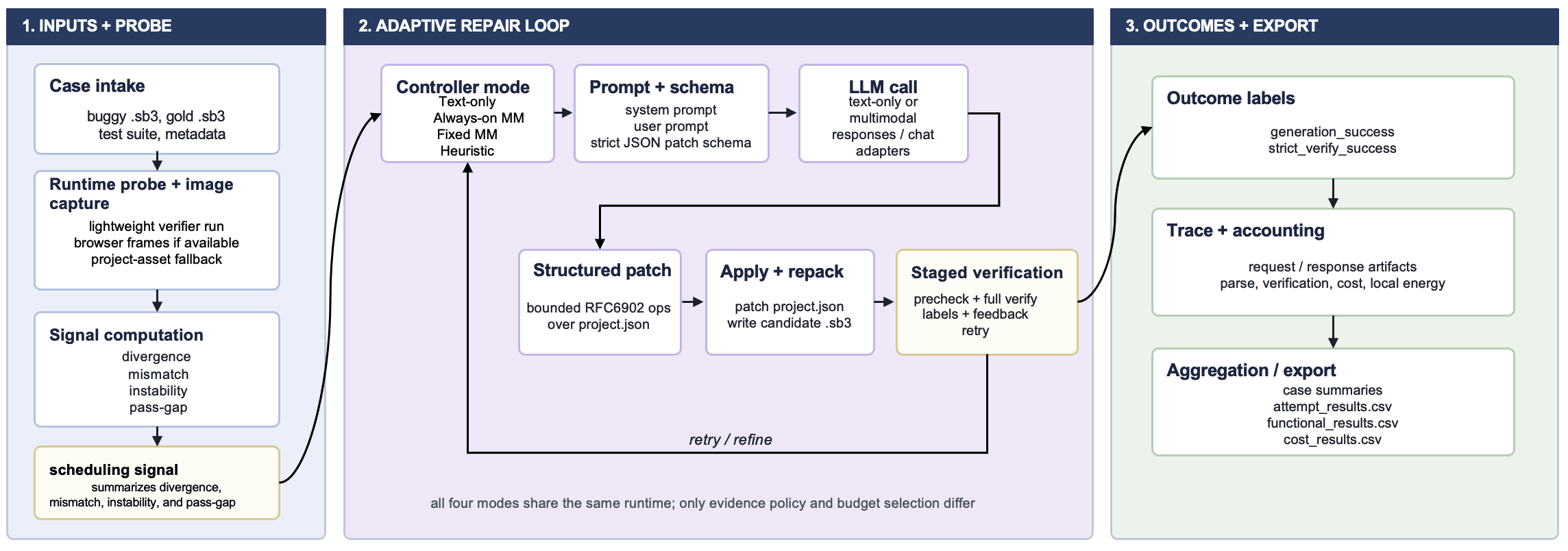}
\caption{\app's end-to-end repair pipeline. Left: the runtime probe executes the buggy project, computes coarse signals, and may capture stage images. Middle: under the selected controller mode in Table~\ref{tab:mode-glossary}, the system maps those signals to a repair plan, builds a bounded JSON-patch request, applies the returned patch, and runs staged verification with retry/refine feedback. Right: the pipeline exports outcome labels and trajectory-level trace, cost, and local-runtime energy records. All four controller modes share the same runtime; only evidence policy and repair-plan selection differ.}
\label{fig:system_pipeline}
\end{figure*}

Table~\ref{tab:mode-glossary} defines the four controller modes referenced throughout this section. Figure~\ref{fig:system_pipeline} is the roadmap for the runtime itself: the left panel shows how a case is reduced to a scheduling signal, the middle panel shows the shared repair loop, and the right panel shows the artifacts exported for evaluation. We follow that same order below: inputs and probe, plan-conditioned repair, then tracing and accounting.

\subsection{Inputs, artifacts, and patch representation}

Each repair case includes a buggy Scratch project (\texttt{.sb3}), a gold reference, an executable test suite, and optional structured metadata. These artifacts keep repair and validation inside one closed loop: the runtime can materialize a concrete candidate and then check whether that candidate moves behavior in the right direction. Because the executable logic lives in the archive's \texttt{project.json}, \app requests structured edits to that representation rather than free-form advice.

Repairs are encoded as bounded RFC 6902-style JSON Patch operations. Each repair plan sets a hard cap on the number of operations, which the response schema enforces. That limit is part of the control policy, not just an output format. It bounds the structural freedom given to the model on a case. The format also keeps edits auditable: each operation names an action and a target path, and the pipeline can distinguish request, extraction, schema, application, and verification failures. In Scratch, many failures come from a single incorrect handler, argument, or broadcast link, so a narrow edit language favors traceable local changes over rewrites.

\subsection{Runtime probe and scheduling signal}

Before any repair attempt, \app runs a lightweight probe through the same black-box verifier interface used later for validation, but with a smaller execution budget. Concretely, the probe executes the buggy project on a short schedule, compares the observed behavior against the reference checks available to the verifier, and records a compact payload that may include buggy pass rate, mismatch rate, rerun instability, the first divergence tick, and optional image artifacts. The probe does not localize or patch the bug. Its job is to summarize whether the failure appears early, broadly, or unreliably, so the controller can decide how much evidence and verification effort the next attempt is likely to need.

The controller then compresses that payload into the coarse scheduling signal shown at the bottom of Figure~\ref{fig:system_pipeline}. Here it is useful to separate \emph{controller mode} from \emph{repair plan}. The controller mode is the outer evaluation condition summarized in Table~\ref{tab:mode-glossary}; the repair plan is the concrete per-attempt choice of evidence policy, JSON Patch budget, and verification schedule. Under Heuristic, the runtime uses the scheduling signal to choose how aggressive the next repair plan should be: cases that diverge early, mismatch broadly, or behave unstably get more evidence and a larger budget, whereas local, stable cases follow a conservative plan. Under Text-only, Always-on multimodal, and Fixed multimodal, the active controller mode fixes the evidence policy and/or repair plan, so the same probe can run without triggering adaptive escalation.

Probe outputs also guide evidence collection. When multimodal execution is enabled by the active controller mode or repair plan, the verifier tries to capture representative stage images during the probe or a short suite run; otherwise the runtime may fall back to project assets. This matters because a controller mode label does not guarantee actual image use. \app therefore records both the images available at runtime and the images actually inserted into the provider payload.

\app does not train a predictor. It relies on a small set of predefined repair plans. Under Heuristic, the scheduling signal selects among those plans; in the three baselines, fixed controller-mode behavior replaces that adaptive mapping. We fix the cutoffs before the matrix run so the evaluation tests the controller design itself rather than a policy tuned after the fact.

\subsection{Prompt construction and multimodal payload assembly}

Once the controller selects a repair plan, \app constructs a system prompt, a user prompt, and a strict JSON schema for the candidate patch. The user prompt includes the signal summary, a compact case description, and prior attempt feedback when available. When localization hints or project excerpts are present, it instructs the model to stay local unless the evidence indicates a broader change.

The adapter layer supports two request styles: a responses-style interface with \texttt{input\_text} and \texttt{input\_image} records, and a chat-style interface with \texttt{image\_url} attachments, including Gemini-compatible routing. For the deployment analysis, the relevant quantity is the realized payload rather than the nominal controller mode label. \app therefore records image counts and byte totals per trajectory so we can separate always-on multimodal execution from selective use.

\subsection{Patch application and staged verification}

The next stages in Figure~\ref{fig:system_pipeline} convert model outputs into executable candidates and then check them under a staged verifier. After the LLM returns, the runtime extracts JSON, validates it against the patch schema, normalizes the operation list, applies the patch to the buggy project's \texttt{project.json}, and repacks the result as a candidate \texttt{.sb3}. If parsing or application fails, the system records the failure and may continue to another attempt.

Verification is staged because full checking is itself part of the cost surface. A repair plan may request a lightweight precheck before the full verifier run, allowing weak candidates to fail early without paying the full cost. Candidates that pass the precheck proceed to the larger run schedule. This is where the paper’s two success notions diverge: generation success requires a successful request, parse, build, and patch application, whereas strict verification success additionally requires passing the strict verifier.

\begin{algorithm}[t]
\caption{Trajectory-level repair loop}
\label{alg:repairloop}
\begin{algorithmic}[1]
\Require case $(b, g, t, m)$ with buggy project $b$, gold reference $g$, test suite $t$, metadata $m$
\State initialize trace writer, cost accumulator, and local energy tracker
\State $p \gets \textsc{Probe}(b, t)$
\State $\sigma \gets \textsc{ComputeSignals}(p)$
\For{$a = 0$ to $\textit{maxAttempts} - 1$}
    \State $\pi \gets \textsc{DecidePlan}(\sigma, a)$
    \State $(s, u, \mathcal{J}) \gets \textsc{BuildPromptAndSchema}(b, \sigma, \pi, m)$
    \State $r \gets \textsc{CallLLM}(s, u, \mathcal{J}, \pi)$
    \State $q \gets \textsc{ParseAndValidate}(r, \pi)$
    \If{$q$ is invalid}
        \State record parse failure and continue
    \EndIf
    \State $c \gets \textsc{ApplyPatch}(b, q)$
    \If{$c$ is invalid}
        \State record apply failure and continue
    \EndIf
    \State $v_1 \gets \textsc{Precheck}(c, g, t, \pi)$
    \If{$v_1$ fails}
        \State record verifier-stage failure and continue
    \EndIf
    \State $v_2 \gets \textsc{FullVerify}(c, g, t, \pi)$
    \State write generation-side and strict-verification-side outcomes
    \If{$v_2$ passes}
        \State \Return exported trajectory summary
    \EndIf
    \State update retry context with verifier feedback
\EndFor
\State stop trackers and export request, parse, verification, cost, and summary artifacts
\end{algorithmic}
\end{algorithm}

Algorithm~\ref{alg:repairloop} expands the middle loop in Figure~\ref{fig:system_pipeline}. It is the allocation unit studied here: probe once, then spend the remaining budget on repair-plan choice, candidate generation, and staged checking. Keeping probe, request, parse, patch, and verification as separate trace stages enables later failure analysis, because unsuccessful runs can be localized and compared across modes.

\subsection{Tracing, cost accounting, and local-runtime energy hooks}

The rightmost panel of Figure~\ref{fig:system_pipeline} corresponds to the artifacts written at the end of each trajectory. The runtime records request and response payloads, parse outputs, verification results, candidate paths, attempt summaries, functional outcomes, and cost records. Run-level aggregation then produces attempt-level, functional, and cost-side summaries together with trajectory-level experimental results. This trace makes the later deployment comparisons auditable rather than anecdotal.

Cost is accumulated per project trajectory rather than per attempt. Energy is tracked at the same level. When available, the implementation uses CodeCarbon-style host-side tracking \cite{codecarbon2022}; otherwise it falls back to a local estimator scoped to the Python process and its child processes. Accordingly, the energy quantity reported in this paper is host-side local-runtime energy: energy for observable host-side work such as orchestration, probing, image capture and preparation, verification, retries, and other pipeline overhead. It excludes provider-side remote inference energy.

\section{Evaluation}
\label{sec:eval}

\subsection{Setup and metrics}

Table~\ref{tab:setup} is the glossary for the evaluation. The unit of analysis is the complete project trajectory, because the controller changes the whole probe--repair--verify loop rather than a single model call. The matrix covers 4,800 trajectories from 12 models and four controller modes, with 100 projects in each model--mode cell. We report results at that level because the controller affects what evidence is collected, how much budget is spent, and how verification proceeds across the trajectory.

The 12 concrete models are Gemini 2.0-flash, 2.5-flash-lite, 2.5-flash, and 2.5-pro, and OpenAI 4.1-nano, 4o-mini, 4.1-mini, 4o, 5-nano, 5-mini, 4.1, and 5.1. Table~\ref{tab:setup} keeps that part of the matrix compact by summarizing the provider split. This trajectory-level accounting follows prior Scratch testing work, which also treats complete executions rather than isolated block fragments as the relevant unit of evidence \cite{stahlbauer2019testing,deiner2023testgen}.

Table~\ref{tab:setup} also defines the recorded quantities. We use \emph{generation success} as the primary deployment metric because the first practical question is whether the controller can carry a case through the loop to an executable candidate, not just how many cases end in a final strict pass. A trajectory counts as generation-successful when the request succeeds, the payload is parsed, and a candidate patch is generated and applied. \emph{Strict verification success} is the main downstream acceptance metric: it counts only when that candidate also passes the strict verifier. Because every strict success in the current matrix is also a generation success, the two metrics separate candidate materialization from downstream acceptance. Cost and host-side local-runtime energy are aggregated per project trajectory and reported in USD/project and Wh/project. The image audit is a supporting measurement: it records input-image count and input-image bytes per trajectory so later sections can distinguish nominal controller modes from actual image use.

\para{Execution environment.} All experiments were run on a local Apple Mac mini (2024) with an Apple M4 chip, 16~GB unified memory, and macOS Tahoe~26.3. The reported local-runtime energy therefore reflects observable host-side work on this machine rather than provider-side inference hardware.

\begin{table}[t]
\caption{Evaluation matrix and recorded quantities used throughout Section~\ref{sec:eval}.}
\label{tab:setup}
\centering
\footnotesize
\renewcommand{\arraystretch}{1.10}
\begin{tabularx}{\columnwidth}{@{}>{\raggedright\arraybackslash}p{0.34\columnwidth}X@{}}
\toprule
\multicolumn{2}{@{}l}{\textbf{Evaluation setup}} \\
Projects & 100 executable Scratch repair projects. \\
Models & 12 concrete models across two provider families (8 OpenAI and 4 Gemini). \\
Modes & Text-only, Always-on multimodal, Fixed multimodal, and Heuristic. \\
Matrix & 4,800 complete project trajectories ($12 \times 4 \times 100$). \\
\midrule
\multicolumn{2}{@{}l}{\textbf{Recorded quantities}} \\
Generation success & The pipeline requests, parses, builds, and applies a candidate patch. \\
Strict verification success & The generated candidate also passes the strict verifier. \\
Cost & Estimated monetary cost, aggregated per project trajectory (USD/project). \\
Image audit & Trajectory-level record of input-image count and input-image bytes; used to interpret actual multimodal use. \\
Energy & Host-side local-runtime energy, aggregated per project trajectory (Wh/project). \\
\bottomrule
\end{tabularx}
\end{table}

\begin{table}[t]
\caption{Overall mode summary across all 4,800 trajectories.}
\label{tab:overall}
\centering
\footnotesize
\renewcommand{\arraystretch}{1.07}
\setlength{\tabcolsep}{3.6pt}
\resizebox{\columnwidth}{!}{
\begin{tabular}{@{}lcccccc@{}}
\toprule
& \multicolumn{2}{c}{Success $\uparrow$} & \multicolumn{4}{c}{Per-trajectory resource use} \\
\cmidrule(lr){2-3}\cmidrule(lr){4-7}
Mode & Gen. & Strict & Cost (\$) & Avg. input images & Attempts & Energy (Wh) \\
\midrule
Text-only & 16.0 & 1.7 & \textbf{0.00148} & \textbf{0.00} & 1.14 & \textbf{0.18} \\
Always-on multimodal & 26.7 & 6.8 & 0.00696 & 2.23 & 1.12 & 0.62 \\
Fixed multimodal & 24.2 & 5.8 & 0.00658 & 2.23 & 1.12 & 0.55 \\
Heuristic & \textbf{30.3} & \textbf{8.0} & 0.00409 & 0.90 & 1.12 & 0.36 \\
\bottomrule
\end{tabular}}
\end{table}

\subsection{Heuristic gives the best overall success--cost--energy tradeoff}

Table~\ref{tab:overall} summarizes the main success--cost--energy tradeoff across the four modes. Text-only is the cheapest and lowest-energy baseline at 0.18~Wh per project, but it also produces the weakest repair outcomes, with 16.0\% generation success and 1.7\% strict verification success. Both non-adaptive multimodal modes improve success, but they do so at substantially higher monetary cost and local-runtime energy. Among the multimodal policies, Heuristic gives the strongest tradeoff: 30.3\% generation success and 8.0\% strict verification success at an average cost of \$0.00409 per project and 0.36~Wh of local-runtime energy.

Relative to Always-on multimodal, Heuristic adds 3.6\pp of generation success, raises strict verification success from 6.8\% to 8.0\%, and cuts average cost and local-runtime energy by 41.2\% and 41.9\%. Relative to Fixed multimodal, it adds 6.1\pp of generation success and 2.2\pp of strict verification while reducing average cost and local-runtime energy by 37.8\% and 34.5\%. The key point is that Heuristic does not buy efficiency by giving up success: within the evaluated budget, it produces more executable candidates and more strict passes while spending less.

\begin{figure*}[t]
    \centering
    \includegraphics[width=\textwidth]{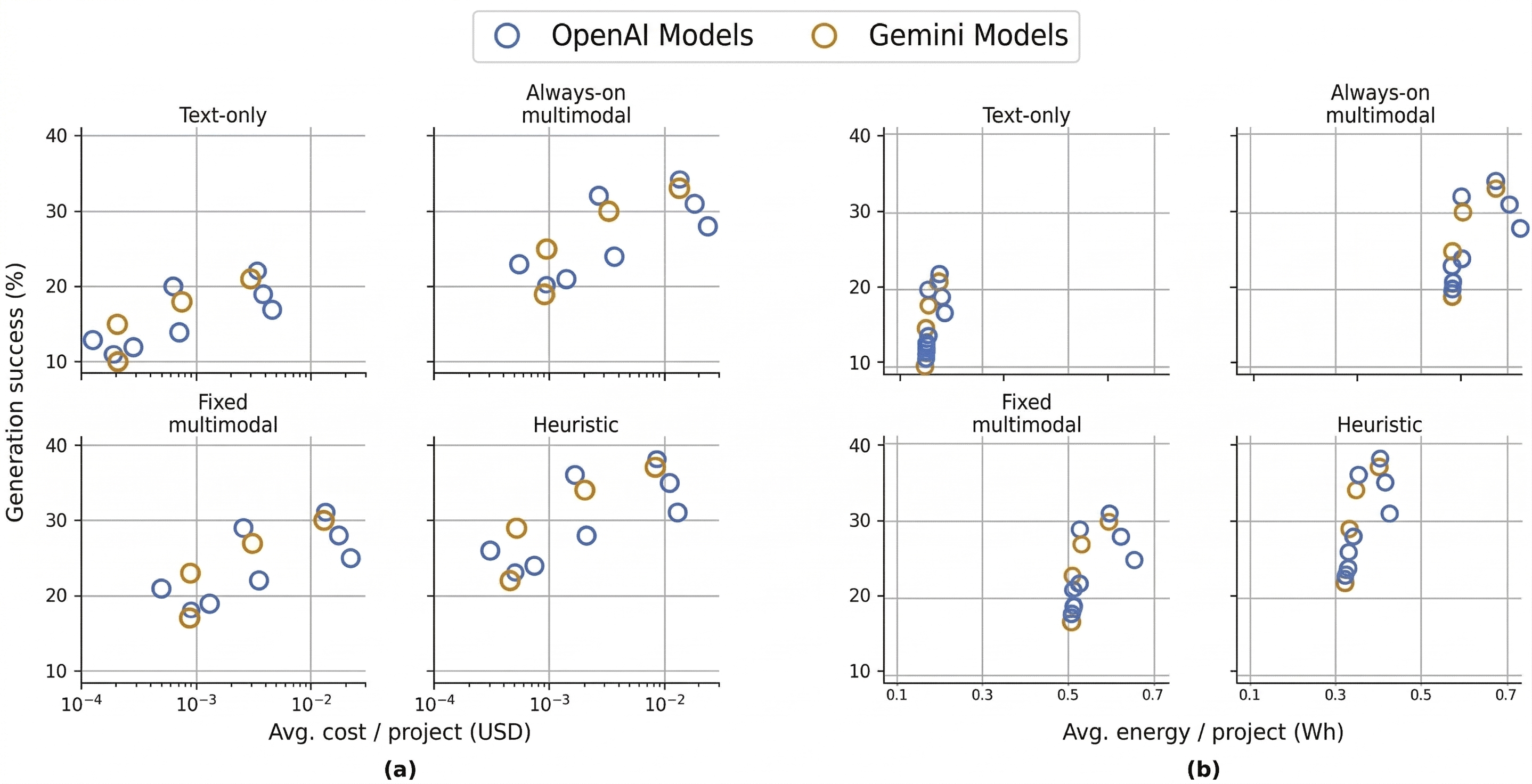}
    \caption{Model-level aggregates for all 48 model--mode cells in the $12 \times 4$ matrix. Each point is one concrete model aggregated over 100 project trajectories. Panel titles encode controller mode; the legend encodes provider family, with blue-outlined hollow points for OpenAI models and orange-outlined hollow points for Gemini models. Panel~(a) plots generation success against average cost per project on a log-scaled x-axis, and panel~(b) plots the same success measure against average local-runtime energy.}
    \label{fig:tradeoff_facets}
\end{figure*}

Figure~\ref{fig:tradeoff_facets} shows that this ranking is not an artifact of pooling. Read each panel as one controller mode: higher points indicate more generation success, and farther-left points indicate lower cost or local-runtime energy. Each point is one model aggregated over 100 project trajectories, and the same ordering appears across the panels. Average attempt counts are nearly identical across Always-on multimodal, Fixed multimodal, and Heuristic: 1.12, 1.12, and 1.12. The gain therefore does not come from giving Heuristic more chances. It comes from using a similar trajectory budget more selectively, avoiding routine image insertion on easier cases while still giving harder cases a larger patch and precheck budget when the probe suggests that visible runtime evidence matters.

\subsection{The Heuristic advantage is consistent across provider families and models}

The next question is whether that ranking survives disaggregation. Table~\ref{tab:family} shows the same ordering within both the OpenAI and Gemini subsets: in each family, Heuristic achieves the highest generation success while remaining clearly cheaper and lower in local-runtime energy than Always-on multimodal and Fixed multimodal. This reduces the plausibility of a provider-specific explanation within the evaluated matrix.

\begin{table}[t]
\caption{Provider-family summary. Generation success is in percent; cost is USD per project; energy is Wh per project.}
\label{tab:family}
\centering
\footnotesize
\renewcommand{\arraystretch}{1.05}
\setlength{\tabcolsep}{5.2pt}
\resizebox{\columnwidth}{!}{
\begin{tabular}{@{}lcccccc@{}}
\toprule
\multirow{2}{*}{Mode} & \multicolumn{3}{c}{OpenAI} & \multicolumn{3}{c}{Gemini} \\
 & Gen. & Cost & Energy & Gen. & Cost & Energy \\
\midrule
Text-only & 16.0 & 0.00170 & 0.18 & 16.0 & 0.00103 & 0.17 \\
Always-on multimodal & 26.6 & 0.00812 & 0.63 & 26.8 & 0.00464 & 0.60 \\
Fixed multimodal & 24.1 & 0.00765 & 0.56 & 24.2 & 0.00442 & 0.54 \\
Heuristic & 30.1 & 0.00473 & 0.36 & 30.5 & 0.00283 & 0.35 \\
\bottomrule
\end{tabular}}
\end{table}

The per-model view is stronger because each model is compared against its own baseline on the same 100 projects. Here the relevant baselines are Always-on multimodal and Fixed multimodal: Table~\ref{tab:overall} already uses Text-only as the minimum-cost floor, whereas this subsection asks whether probe-guided multimodal scheduling improves on non-adaptive multimodal policies. Table~\ref{tab:dominance} counts pairwise wins. Heuristic beats Always-on multimodal on generation success and cost for all 12 models, and on strict verification for 11 models while tying the remaining one. Against Fixed multimodal, it wins on generation success, cost, and strict verification for all 12 models. Energy follows the same pattern: for every model, Heuristic uses less average local-runtime energy than both non-adaptive multimodal baselines, with reductions of 40.2\%--43.8\% relative to Always-on multimodal and 32.4\%--36.6\% relative to Fixed multimodal.

\begin{table}[t]
\caption{Pairwise dominance of Heuristic over the two non-adaptive multimodal baselines. Text-only is omitted because this table isolates adaptive versus non-adaptive multimodal control.}
\label{tab:dominance}
\centering
\footnotesize
\setlength{\tabcolsep}{3.0pt}
\renewcommand{\arraystretch}{1.05}
\begin{tabularx}{\columnwidth}{@{}Ycccc@{}}
\toprule
Comparison & Gen. wins & Strict wins & Cost wins & Gain / cost range \\
\midrule
\makecell[l]{Heuristic vs.\\Always-on\\multimodal} & 12/12 & 11/12 & 12/12 & \makecell[l]{+3.0 to +4.0\pp;\\37.1 to 50.2\%} \\ \midrule
\makecell[l]{Heuristic vs.\\Fixed\\multimodal} & 12/12 & 12/12 & 12/12 & \makecell[l]{+5.0 to +7.0\pp;\\34.0 to 48.5\%} \\
\bottomrule
\end{tabularx}
\end{table}

\begin{figure}[t]
    \centering
    \includegraphics[width=.47\textwidth]{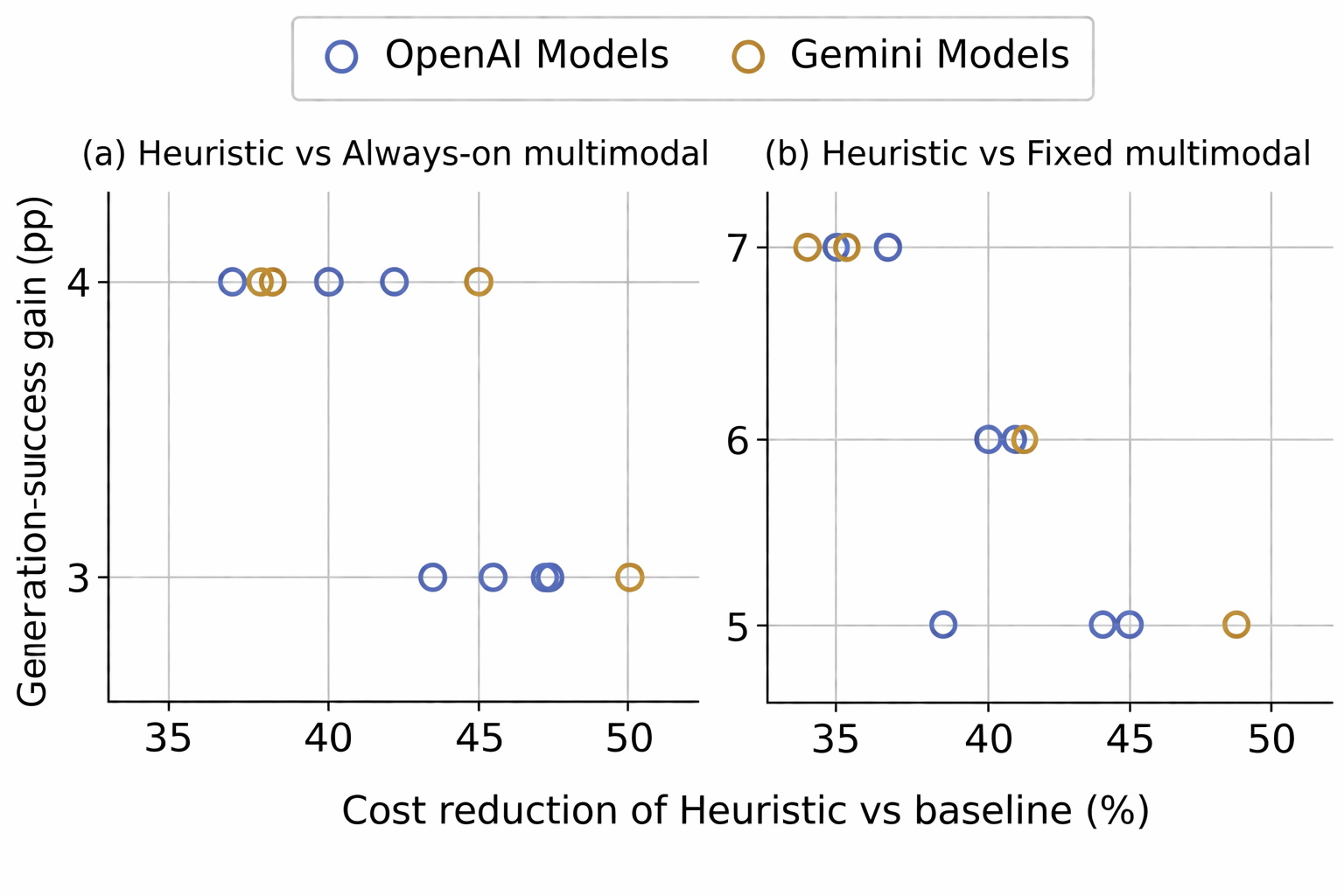}
    \caption{Per-model deltas of Heuristic relative to the two non-adaptive multimodal baselines. Each point denotes one concrete model aggregated over 100 project trajectories; the legend encodes provider family, with blue-outlined hollow points for OpenAI models and orange-outlined hollow points for Gemini models. Panel~(a) uses Always-on multimodal as the baseline and panel~(b) uses Fixed multimodal. The x-axis is cost reduction (\%), the y-axis is generation-success gain (pp), and all points lie in the positive-positive quadrant.}
    \label{fig:deltas}
\end{figure}

Table~\ref{tab:modeldeltas} and Figure~\ref{fig:deltas} show why those aggregate wins matter. The gains are modest but consistent, not driven by a small number of outliers: generation success rises by 3--4\pp relative to Always-on multimodal and by 5--7\pp relative to Fixed multimodal, while cost and local-runtime energy fall for every model. Because the comparison here isolates adaptive versus non-adaptive multimodal policies, there is intentionally no Heuristic-versus-Text-only panel. Figure~\ref{fig:deltas} is best read as a consistency map: every point in the positive-positive quadrant marks a model where Heuristic improves generation success while lowering cost.

\begin{table}[t]
\caption{Per-model deltas of Heuristic relative to the same two non-adaptive multimodal baselines used in Figure~\ref{fig:deltas}. Generation gains are percentage-point differences; cost and local-runtime energy columns report relative reductions (\%). Text-only is omitted because Table~\ref{tab:overall} already serves as the minimum-cost floor.}
\label{tab:modeldeltas}
\centering
\footnotesize
\renewcommand{\arraystretch}{1.08}
\setlength{\tabcolsep}{3pt}
\resizebox{\columnwidth}{!}{
\begin{tabular}{@{}l@{\hspace{-6pt}}rrrrrr@{}}
\toprule
& \multicolumn{3}{c}{vs. Always-on multimodal} & \multicolumn{3}{c}{vs. Fixed multimodal} \\
\cmidrule(lr){2-4}\cmidrule(lr){5-7}
Model & Gen. gain (pp) & Cost red. (\%) & Energy red. (\%) & Gen. gain (pp) & Cost red. (\%) & Energy red. (\%) \\
\midrule
\multicolumn{7}{@{}l}{\textit{Gemini}} \\
\texttt{2.0-flash} & +3.0 & 50.2 & 43.6 & +5.0 & 48.5 & 36.6 \\
\texttt{2.5-flash-lite} & +4.0 & 44.9 & 41.8 & +6.0 & 41.2 & 34.7 \\
\texttt{2.5-flash} & +4.0 & 38.4 & 41.9 & +7.0 & 34.0 & 34.7 \\
\texttt{2.5-pro} & +4.0 & 38.0 & 40.4 & +7.0 & 35.4 & 32.6 \\
\addlinespace[2pt]
\multicolumn{7}{@{}l}{\textit{OpenAI}} \\
\texttt{4.1-nano} & +3.0 & 47.3 & 43.8 & +5.0 & 44.8 & 36.6 \\
\texttt{4o-mini} & +3.0 & 47.4 & 42.8 & +5.0 & 44.0 & 35.9 \\
\texttt{4.1-mini} & +4.0 & 42.2 & 42.6 & +6.0 & 40.0 & 35.1 \\
\texttt{4o} & +3.0 & 45.5 & 42.6 & +6.0 & 40.9 & 35.0 \\
\texttt{5-nano} & +3.0 & 43.5 & 42.3 & +5.0 & 38.5 & 35.3 \\
\texttt{5-mini} & +4.0 & 38.0 & 40.7 & +7.0 & 35.0 & 33.0 \\
\texttt{4.1} & +4.0 & 40.1 & 41.2 & +7.0 & 36.5 & 33.5 \\
\texttt{5.1} & +4.0 & 37.1 & 40.2 & +7.0 & 35.3 & 32.4 \\
\bottomrule
\end{tabular}}
\end{table}

\subsection{Heuristic stays genuinely multimodal while preserving strict-verification quality}

Section~\ref{sec:eval} leaves two follow-up questions after the aggregate tradeoff results: does Heuristic win only by drifting back toward text-only behavior, and do its extra generation successes survive the stricter verifier? Table~\ref{tab:imageaudit} addresses both. Text-only never uses images, whereas Always-on multimodal and Fixed multimodal both use them on 78.0\% of project trajectories. Heuristic still uses images on 31.3\% of trajectories. It is therefore a genuinely multimodal policy, but a selective one rather than a uniform one.

\begin{table}[t]
\caption{Image-use and strict-verification audit by mode. The first three columns show how many generation-success trajectories also become strict successes; the last column shows how often a mode actually uses images. The strict ratio is computed over generation-success trajectories.}
\label{tab:imageaudit}
\centering
\footnotesize
\renewcommand{\arraystretch}{1.05}
\setlength{\tabcolsep}{4pt}
\resizebox{\columnwidth}{!}{
\begin{tabular}{@{}lcccc@{}}
\toprule
Mode & Gen. count & Strict count & Strict/gen. (\%) & Trajectories using images (\%) \\
\midrule
Text-only & 192 & 20 & 10.4 & 0.0 \\
Always-on multimodal & 320 & 81 & 25.3 & 78.0 \\
Fixed multimodal & 290 & 69 & 23.8 & 78.0 \\
Heuristic & 363 & 96 & 26.4 & 31.3 \\
\bottomrule
\end{tabular}}
\end{table}

Table~\ref{tab:imageaudit} also clarifies the strict-verification result. The multimodal modes convert roughly one quarter of generated candidates into strict-verifier passes, and Heuristic does not obtain its gains by producing weaker candidates: its strict-to-generation ratio is 26.4\%, slightly above Always-on multimodal (25.3\%) and Fixed multimodal (23.8\%). In other words, selective escalation increases the number of executable candidates without reducing the share that survives the stricter execution check. If we looked only at final strict passes, we would miss part of the story: even the best mode records 96 strict successes out of 363 generated candidates, so strict verification alone would hide meaningful gains in candidate production and application.

\subsection{Failure-layer observations}

The recorded traces also show where unsuccessful trajectories stop, which helps explain the mode-level differences above. A trajectory can fail before candidate application because the response cannot be parsed or because the patch cannot be applied, or it can survive long enough to fail at the verifier stage after candidate application. Under the two non-adaptive multimodal modes, many trajectories stop in those earlier layers. Always-on multimodal records 463 parse failures and 277 application failures, whereas Fixed multimodal records fewer parse failures (293) but more application failures (472). Heuristic reduces parse failures to 279 and keeps application failures below Fixed multimodal at 412.

As a result, more unsuccessful trajectories reach the verifier stage under Heuristic: 267 verifier-stage failures after candidate application, compared with 239 for Always-on multimodal and 221 for Fixed multimodal. One plausible reason is payload complexity: richer multimodal inputs and larger repair budgets can increase malformed JSON or poorly grounded patches before verification. Heuristic does not remove those earlier failures, but it shifts more unsuccessful trajectories into verifier-stage failure. Operationally, that is preferable because those trajectories still produce an applied candidate and a more informative trace.

\section{Discussion}
\label{sec:discussion}

\para{Why the heuristic controller wins.}
The heuristic controller’s main advantage is allocation rather than additional search. Its average attempt count is essentially unchanged from the Fixed multimodal baseline, so improvements come not from extra retries but from when multimodal effort is spent. Image use drops sharply relative to uniform policies yet remains substantial, avoiding routine overspending without collapsing into text-only repair. The same ordering holds across all 12 models and both provider families: Heuristic improves generation success, lowers cost, and reduces local-runtime energy for every model, which is more consistent with better budget allocation than with a provider-specific effect. Failure traces reinforce this interpretation. Compared with Always-on and Fixed multimodal policies, the heuristic controller loses fewer trajectories to early parse or application failures and sends more to the verifier stage, a shift that matters because later failures still yield an applied candidate and a diagnostic trace, whereas malformed outputs consume budget without useful feedback.

\para{Selective multimodality as an operational policy.}
The image audit makes the policy concrete. Uniform multimodal strategies assume visual evidence is usually worth sending, yet Heuristic performs better while using images on only 31.3\% of trajectories, compared with 78.0\% under Fixed multimodal. The implication is that image use should follow runtime evidence and be paid for when stage-level behavior is likely to matter. More broadly, artifact collection should be treated as a selective design choice rather than default preprocessing. Treating every case as evidence-intensive can waste resources even with strong models. This view complements learner-facing systems such as iSnap, tutorial feedback pipelines, and Stitch \cite{marwan2023isnap,obermueller2023tutorials,si2025stitch}, which aim to support human learning, whereas \app is evaluated here as autonomous repair infrastructure for novice-authored projects, emphasizing budget allocation and verifier behavior rather than pedagogy.

\para{Implications for deployment design.}
These results suggest that better multimodal repair does not necessarily require a stronger model or a more elaborate prompt. The heuristic controller studied here is intentionally simple: it uses a lightweight probe, a small signal set, and threshold-based plans, yet Heuristic consistently outperforms uniform multimodal policies on success, cost, and local-runtime energy. In this matrix, evidence allocation looks at least as consequential as model choice for the observed tradeoffs.

That gain is realized across the whole repair trajectory, not at a single prompt exchange. This is not an argument against multimodality. Images remain useful precisely when text-only repair is most likely to miss important runtime behavior. For deployment, the practical lesson is to treat artifact collection as resource allocation: gather richer evidence when preliminary signals justify it, and defer it otherwise.

\para{Why generation success and strict verification capture different aspects of pipeline quality.}
Generation success and strict verification mark different cut points in the same pipeline. Generation success measures whether the system produces a valid, applicable patch. Strict verification asks whether that patch survives a stronger execution-based test. Lower strict verification rates are therefore expected, and keeping both metrics makes the deployment picture clearer. A system that produces more valid candidates can still be valuable even when some of those candidates fail stricter checks, because candidate materialization and final strict acceptance capture different operational stages of the pipeline. Our results show that the heuristic controller improves both metrics without implying guaranteed correctness.

\para{What the energy numbers mean.}
The energy numbers should be read in the same operational sense as the cost numbers: they report one local-runtime energy measurement per full project trajectory, aggregating observable host-side work over the whole repair loop. On that measure, Text-only is lowest at 0.18~Wh per project. Among multimodal modes, Heuristic uses substantially less energy than Always-on and Fixed policies---0.36~Wh versus 0.62~Wh and 0.55~Wh per project, respectively. Because each trajectory has only one aggregate measurement, we cannot attribute that reduction to individual stages. The pattern is consistent with less image processing and less unnecessary pipeline work, but the data do not support a finer decomposition. The measurements also exclude provider-side inference energy, so the claim here is comparative local deployment efficiency, not total system energy use.

\section{Threats to Validity}
\label{sec:threats}

\para{Internal Validity.}
The runtime probe measures observable behavioral discrepancy rather than true repair complexity, so it only approximates downstream difficulty. Divergence, mismatch, instability, and buggy pass gap are useful scheduling signals but do not capture full fault structure: apparently easy cases may require large edits, while severe ones may admit concise local fixes. We evaluate the heuristic controller under a single configuration, so different thresholds or budgets could shift the capability cost tradeoff, especially near decision boundaries where small probe changes flip the plan. The pipeline is also stochastic, with provider outputs, parsing, and verification varying across executions. The matrix therefore records one realized trajectory per project, mode, and model rather than repeated runs with confidence intervals, and we make no statistical significance claims, relying instead on the fully crossed paired design and the consistency of patterns across all 12 models. A separate threat concerns result reconstruction. Aggregates are regenerated automatically from completed results, which removes transcription error but introduces risks from joins, deduplication, path alignment, or derived fields such as energy. We mitigate these risks by verifying that every model-mode cell contains exactly 100 projects, confirming coverage of all 4,800 trajectories, and recomputing aggregates from raw trajectory records. Reported validity is thus bounded by the observed matrix rather than statistical generalization: the paired design stabilizes within model comparisons but cannot eliminate single run variance or provider drift, so small cross-mode gaps should be read as directional unless they recur consistently at the per-model level.

\para{External Validity.}
Our 100-project executable corpus covers a broad but still incomplete slice of Scratch repair tasks. It supports controlled comparison, but not the full diversity of classroom workflows, open-ended creative projects, or forum-based help requests. The benchmark also favors cases that can be converted into stable executable repairs with clear test oracles, so it may underrepresent problems that depend on subjective or context-sensitive judgment. The multimodal evidence path is narrower than in many real deployments: the current pipeline mainly uses probe-captured images and fallback project assets, whereas classroom debugging may include longer gameplay traces, teacher annotations, or interactive learner feedback. A controller tuned to this evidence channel may therefore need recalibration when the available runtime context becomes broader or noisier. The model set is also time-bounded. Provider behavior, pricing, modality support, and routing policies change quickly, so absolute costs and some rankings may shift. We therefore emphasize relative control effects within the matrix rather than durable provider comparisons. Local-runtime energy is likewise environment-sensitive and depends on hardware, instrumentation, and system load. Finally, the study evaluates autonomous repair rather than classroom use, so it does not establish pedagogical effectiveness, learner trust, or usability in real teaching settings.

\para{Construct Validity.}
The primary constructs are operational by design. Generation success asks whether the pipeline produces and applies a candidate, not whether the repair is ultimately correct. Strict verification success is a stronger but narrower construct defined by the current verifier. Reporting both separates candidate delivery from final acceptance, but neither should be interpreted as a complete measure of repair quality.

Image use is quantified through trajectory-level counts and bytes rather than request-level semantic attribution, and average cost is computed per full project trajectory, the deployment unit of interest here. This aggregation compresses tail behavior that may matter for latency-sensitive settings, although failure-layer analysis and per-model breakdowns partly expose such effects. Energy reflects observable host-side runtime work only; it excludes remote inference, datacenter consumption, and fine-grained hardware breakdowns, and the merged project-trajectory results contain no separate duration or emissions fields. The results should therefore be interpreted under these operational definitions.

\section{Related Work}
\label{sec:related}

\para{Scratch debugging and automated feedback.}
Scratch already has a broad support ecosystem spanning static analysis, hinting, debugging, logging, and automated feedback. Static and semi-static tools include Dr.~Scratch, which evaluates computational-thinking properties and common bad habits \cite{morenoleon2015drscratch}; LitterBox, which lint-checks Scratch programs \cite{fraser2021litterbox}; and the common-bug catalog, which characterizes recurrent bug patterns and code smells \cite{fraedrich2021commonbugs}. These systems surface likely issues early, but they do not attempt autonomous repair of open-ended bugs under deployment constraints.

Another line of work studies learner-facing hints and feedback. Catnip generates next-step hints from automated tests \cite{obermueller2021catnip}. iSnap studies data-driven hinting for block-based programming \cite{marwan2023isnap}. Adaptive Immediate Feedback examines real-time feedback timing and granularity in block-based tasks \cite{marwan2022aif}. Effects of Hints on Debugging Scratch Programs studies how hint quality changes debugging effort \cite{greifenstein2021hints}, and Effects of Automated Feedback in Scratch Programming Tutorials evaluates automated feedback in tutorial settings \cite{obermueller2023tutorials}. More recently, Debugging with an AI Tutor analyzes how novices seek and interpret conversational debugging help \cite{yang2024aitutor}, while Stitch moves toward explanation-first, reference-guided tutoring through discrepancy analysis and stepwise guidance \cite{si2025stitch}. These systems are close in domain, but their goal is pedagogical support rather than autonomous repair-time control.

A third line of work focuses on debugger-style observability. NuzzleBug provides stepping, breakpoints, reverse execution, and interrogative debugging for Scratch \cite{deiner2024nuzzebug}, while ScratchLog records live learner activity for analytics \cite{caspari2023scratchlog}. Both reinforce the importance of runtime evidence in Scratch. \app builds on the same observation, but uses runtime signals to schedule multimodal repair rather than to support tutoring, analytics, or direct user-driven debugging.

\para{Executable Scratch benchmarks and infrastructure.}
Rigorous Scratch evaluation depends on executable infrastructure. Whisker demonstrates automated and property-based testing and shows that VM-level execution can replace manual checking in many scenarios \cite{stahlbauer2019testing}. Subsequent work automates test generation for Scratch \cite{deiner2023testgen} and integrates block-based testing directly into Scratch \cite{feldmeier2024blocktesting}. More recent efforts package curated projects, injected bugs, and executable oracles into reproducible artifacts, as in ScratchEval \cite{si2026scratcheval}.
\app follows this executable, artifact-driven evaluation paradigm, but it does not introduce a new benchmark or testing technique. Instead, it studies a deployment question on top of that substrate: when is additional evidence worth its cost during multimodal repair? Stable assets, harnesses, and executable oracles make paired trajectory comparisons possible, which lets \app focus on repair-time control rather than benchmark construction.

\para{Multimodal Scratch reasoning.}
Recent Scratch work increasingly treats runtime visuals as evidence rather than cosmetic context. ViScratch shows that gameplay video can materially improve diagnosis and repair for Scratch bugs that are difficult to localize from blocks alone \cite{si2025viscratch}. MindScratch extends multimodal generative-AI support to classroom-oriented visual programming assistance \cite{chen2025mindscratch}. Stitch connects LLM-based guidance to discrepancy-aware tutoring in Scratch \cite{si2025stitch}, and recent work on AI-tutor-mediated debugging further examines how novices interact with conversational support during debugging \cite{yang2024aitutor}. ScratchEval packages multimodal executable cases for systematic evaluation \cite{si2026scratcheval}, while NuzzleBug shows why runtime observability remains central even outside explicit multimodal prompting \cite{deiner2024nuzzebug}.
\app departs from this literature at the control layer. We do not revisit whether visual evidence can help. Instead, we ask how a deployed repair pipeline should schedule multimodality, patch budget, and verification effort after a lightweight runtime probe. 

\para{Cost-aware and energy-aware AI systems.}
Green AI argues that capability claims should be paired with efficiency reporting and disciplined measurement \cite{schwartz2020green,zhang2025systematicstudytimelimit}. Earlier work on deep learning in NLP likewise emphasized the financial and environmental cost of large-model development \cite{strubell2019energy}. Tooling such as CodeCarbon has made host-side accounting easier to integrate into practical ML workflows \cite{codecarbon2022}. More recent studies revisit the environmental footprint of modern LLMs, including comparative assessments of energy, water, carbon, and cost \cite{ren2024reconciling} and infrastructure-aware benchmarking of LLM inference footprints \cite{jegham2025hungry}.
\app adopts that stance with a deliberately narrow scope. The system records probe work, request artifacts, verification effort, and trajectory-level monetary cost in a unified trace, and it can attach local-runtime energy estimates derived from host-side tracking \cite{codecarbon2022}. It does not estimate provider-side hardware energy. The goal is to analyze operational control inside an executable Scratch repair loop, not to make a full life-cycle claim about commercial model providers.

\para{Verifier-guided and iterative program repair.}
Automated repair has long separated generation from validation. GenProg is a canonical generate-and-validate system \cite{legoues2012genprog}, and later surveys make clear that execution-based feedback remains central to repair pipelines \cite{gazzola2019asr,clef,9825801}. Recent LLM-centered work \cite{pydex,gmerge} preserves the same structure at a different scale. SWE-bench couples model outputs with executable issue resolution \cite{jimenez2024swebench}, while SWE-agent makes the environment loop more explicit \cite{yang2024sweagent}.
\app fits this lineage, but the cost surface differs. Scratch is block-based, event-driven, and partly visual, so candidate quality depends not only on textual reasoning but also on probe runs, image capture, patch materialization, and staged verification. Our focus is the control policy over this loop rather than a general-purpose repair agent.

\section{Conclusion}
\label{sec:conclusion}

\app treats multimodal Scratch repair as a deployment-control problem for executable projects: the key question is not just how to generate a fix, but when richer evidence and extra verification are worth the cost. The system probes each project, uses that signal to decide whether to stay text-only or escalate, and keeps the resulting trajectory auditable through bounded patches and staged verification. Across the fully crossed matrix of 12 models, four modes, and 100 projects, Heuristic delivers the strongest observed multimodal tradeoff. It improves on the two non-adaptive multimodal baselines while reducing average monetary cost and local-runtime energy, whereas Text-only remains the absolute energy floor. Within the evaluated bounded trajectory budget, multimodal evidence works best when it governs escalation rather than being applied by default. In an end-to-end repair loop, controller choice can matter as much as model choice.

\bibliography{scratch}

\end{document}